\documentclass[11pt, secnumarabic,amssymb, nobibnotes, aps, prd]{revtex4-2}
	
\usepackage[dvips]{graphicx,color}
\usepackage{times}
\usepackage{filecontents}
\usepackage{mathtools}
\usepackage{subcaption}
\usepackage{graphicx}
\usepackage{amsmath}
\usepackage{multirow}
\usepackage{xcolor}

\usepackage[%
  colorlinks=true,
  urlcolor=blue,
  linkcolor=red,
  citecolor=blue
]{hyperref}

\begin{document}
\title{Lyapunov Exponents and Phase Transition of Hayward AdS Black Hole}

\author{Naba Jyoti Gogoi$^1$  }\email{gogoin799@gmail.com }
\author{Saumen Acharjee$^1$} \email{saumenacharjee@dibru.ac.in}
\author{Prabwal Phukon$^{1,2}$} \email{prabwal@dibru.ac.in}

\affiliation{$1.$ Department of Physics, Dibrugarh University,
Dibrugarh 786004, Assam, India \\
$2.$Theoretical Physics Division, Centre for atmospheric studies, Dibrugarh University}

\begin{abstract}
In this paper, we study the relationship between the phase transition and Lyapunov exponents for 4D Hayward anti-de Sitter (AdS) black hole. We consider the motion of massless and massive particles around an unstable circular orbit of the Hayward AdS black hole in the equatorial plane and calculate the corresponding Lyapunov exponents. The phase transition is found to be well described by the multivaled Lyapunov exponents. It is also found that different phases of Hayward AdS black hole coincide with different branches of the Lyapunov exponents. We also study the discontinuous change in the Lyapunov exponents and find that it can serve as an order parameter near the critical point. The critical exponent of change in Lyapunov exponent near the critical point is found to be $1/2$.
\end{abstract}


\maketitle
\section{Introduction}
Black hole thermodynamics \cite{Hawking:1971tu,Bekenstein:1973ur,Hawking:1974rv,Hawking:1975vcx,Bardeen:1973gs}   has its roots in the similarities between black hole mechanics and thermodynamics. Over the last few decades, it has emerged as a vibrant and dynamic field of study, drawing considerable attention from researchers. Over time, numerous remarkable investigations have unveiled a plethora of fascinating and thought-provoking findings \cite{Wald:1979zz,bekenstein1980black,Wald:1999vt,Carlip:2014pma,Wall:2018ydq,Candelas:1977zz,Chamblin:1999hg,Hawking:1982dh,Chamblin:1999tk}. The introduction of the AdS/CFT correspondence \cite{Maldacena:1997re,Gubser:1998bc} has motivated researchers for an extensive exploration into the thermodynamics and phase structure for a number of AdS black holes \cite{Kubiznak:2012wp,Altamirano:2013ane,Altamirano:2013uqa,Wei:2014hba,Frassino:2014pha,Cai:2013qga,Xu:2014tja,Dolan:2014vba,Hennigar:2015esa,Hennigar:2015wxa,Hennigar:2016xwd,Zou:2016sab,Gogoi:2021syo,Gogoi:2023qni}. Different approaches such as Ruppeiner geometry \cite{Ruppeiner:2012uc,Miao:2017cyt,Guo:2019oad,Wei:2019yvs,Wang:2019cax,Yerra:2020oph,Yerra:2021hnh}, thermodynamic topology \cite{Wu:2022whe,Liu:2022aqt,Fan:2022bsq,Gogoi:2023qku,Gogoi:2023xzy,Ye:2023gmk,Zhang:2023uay,Gogoi:2023wih,Du:2023wwg,Fairoos:2023jvw,Du:2023nkr,Wu:2023sue,Wu:2023xpq,Ali:2023zww,Sadeghi:2023dsg,Saleem:2023oue,Shahzad:2023cis,Chen:2023elp,Bai:2022klw,Yerra:2022alz,Hazarika:2023iwp,Mehmood:2023psa,Tong:2023lyc,Wang:2023qxw,Sadeghi:2023tuj,Wu:2023fcw,Wu:2023meo} etc. have been use to study phase transition from different perspectives. 

Of particular interest, endeavors have been made to link the phase transitions of black holes with observable phenomena. These efforts have explored potential connections between black hole phase transitions and various observational signatures, including characteristics such as quasinormal modes \cite{Liu:2014gvf,Zou:2017juz,Zhang:2020khz,Mahapatra:2016dae,Chabab:2016cem}, the circular orbit radius of test particles \cite{Wei:2017mwc,Wei:2018aqm,Zhang:2019tzi}, and the radius of the black hole shadow \cite{Zhang:2019glo,Belhaj:2020nqy}.

One fascinating phenomenon worth mentioning is the motion of particles around black holes because they can provide some important information regarding the background spacetimes. For instance, it has been found that the unstable circular null geodesics can impact the optical appearance of a star experiencing gravitational collapse, potentially elucidating the exponential fade-out in luminosity observed during the collapse process \cite{Ames}. Also, the null geodesics are found to be useful in explaining the quasinormal modes (QNMs) of a black hole \cite{Cardoso:2008bp,Konoplya:2017wot,Konoplya:2019hml}. Such motion of particles may cause a very important phenomenon in non-linearly dynamic systems known as chaos and to study a chaotic system, Lyapunov exponents can be used \cite{SA_1}. It  provides a straightforward means of characterizing the dynamics of a chaotic system by examining its effective degrees of freedom. Extensive research has been conducted on the chaotic motion of particles in black hole spacetime \cite{Sota:1995ms,Kan:2021blg,Gwak:2022xje,Hanan:2006uf,Gair:2007kr,AlZahrani:2013sqs,Polcar:2019kwu,Wang:2016wcj,Chen:2016tmr,Lu:2018mpr,Guo:2020xnf}.  A universal upper bound for the Lyapunov exponents in thermal quantum systems is found in \cite{Maldacena:2015waa}. 
Nevertheless, it is found to be violated in some cases studied in \cite{Zhao:2018wkl,Guo:2020pgq}.

A recent conjecture proposed in \cite{Guo:2022kio} suggests a relationship between Lyapunov exponents and the phase transition of black holes. The work shows that the Lyapunov exponents become multivalued during phase transition and become single valued when there is no phase transition. Furthermore, it is found that the discontinuous change in the Lyapunov exponents can be treated as an order parameter, yielding a critical exponent of $1/2$ near the critical point. This conjecture has been further verified for different black holes \cite{Lyu:2023sih,Yang:2023hci,Kumara:2024obd,Du:2024uhd}.

In this work, we extend the study of Lyapunov exponents to a regular black hole named Hayward AdS black hole which was first proposed by Sean A. Hayward \cite{Hayward:2005gi}. This black hole differs notably from conventional black holes like the Schwarzschild and Reissner-Nordstr\"om black holes, which typically exhibit singularities at their centres. The thermodynamics and phase transition of Hayward black hole in AdS spacetime have been extensively studied in a number of remarkable works \cite{Mehdipour:2016rtd,NaveenaKumara:2020lgq,Fan:2016rih}. We study the Lyapunov exponent of massless and massive particles in an unstable circular orbit in the equatorial plane around the Hayward AdS black hole and study its relationship with the phase transition. We also investigate the behavior of the Lyapunov exponents near the critical point and calculate the critical exponent.

This paper is organized as follows: In \autoref{Thermodynamics}, we review the thermodynamics and phase structure of the Hayward AdS black hole. In \autoref{LyapunovExponent}, the Lyapunov exponents for massless and massive particles are discussed in their respective subsections. In this section, we also study the relationship between Lyapunov exponents and the phase structure of Hayward AdS black hole. Finally, we conclude our results in \autoref{Conclusion}.

\section{Thermodynamics and phase structure of Hayward AdS black hole}
\label{Thermodynamics}
The static and spherically symmetric Hayward AdS black hole is represented by the following line element \cite{Fan:2016rih}:
\begin{equation}
ds^2=-f(r)dt^2+\frac{dr^2}{f(r)}+r^2 ( d\theta^2 + \sin\theta^2 d\phi^2) \quad \text{with}  \quad A=Q_m\cos\theta d\phi,
\end{equation}
and
\begin{equation}
\label{Eq:Fr}
f(r)=1+\frac{r^2}{l^2}-\frac{2 M r^2}{r^3+q^3},
\end{equation}
where $M$ is the mass of the black hole, $l$ is the AdS length and $q$ is an integration constant which is related to the total magnetic charge $Q_m$ of the black hole by
\begin{equation}
Q_m=\frac{q^2}{\sqrt{2\alpha}},
\end{equation}
in which, $\alpha$ is a parameter associated with the non-linear electromagnetic field. The mass of the black hole $M$ can be calculated by the condition $f(r_+)=0$, which yields
\begin{equation}
\label{Eq:Mass}
M=\frac{\left(l^2+r_+^2\right) \left(q^3+r_+^3\right)}{2 l^2 r_+^2}.
\end{equation}
Here, $r_+$ is the horizon radius of the black hole.
The Hawking temperature $T$ and entropy $S$ are given by 
\begin{equation}
\label{Eq:HawTem}
T=\frac{l^2 \left(r_+^3-2 q^3\right)+3 r_+^5}{4 \pi  l^2 r_+ \left(q^3+r_+^3\right)},
\end{equation} 
and
\begin{equation}
S=2 \pi  \left(\frac{r_+^2}{2}-\frac{q^3}{r_+}\right).
\end{equation}
The first law of thermodynamics has the following form
\begin{equation}
dM=TdS+\Psi dQ_m+VdP+\Pi d\alpha,
\end{equation}
where $\Psi$ and $\Pi$ are conjugate parameters corresponding to $Q_m$ and $\alpha$. $P$ is the pressure with its conjugate volume $V$.
The Gibbs free energy can be calculated from the definition as
\begin{equation}
\label{Eq:Gibb_Free_Energy}
F=M-TS=\frac{l^2 \left(8 q^3 r_+^3-2 q^6+r_+^6\right)+10 q^3 r_+^5+2 q^6 r_+^2-r_+^8}{4 l^2 r_+^2 \left(q^3+r_+^3\right)}.
\end{equation}
Now, we use dimensional analysis and scale the following quantities as 
\begin{equation}
\label{Eq:AdsScaling}
\tilde{r}_+=r_+/l, \quad \tilde{q}=q/l , \quad \tilde{F}=F/l , \quad  \tilde{T}=T l , \quad \text{and}  \quad \tilde{M}=M/l.
\end{equation}
The tilde symbol is used to denote dimensionless quantities. 

The critical points can be calculated by using the condition 
\begin{equation}
\frac{\partial\tilde{T}}{\partial \tilde{r}_+}=\frac{\partial^2\tilde{T}}{\partial^2 \tilde{r}_+}=0,
\end{equation}
where, we have used the Hawking temperature \eqref{Eq:HawTem}  along with the scaling \eqref{Eq:AdsScaling}. There exists a single critical point for each corresponding quantity and their numerical values are given by 
\begin{equation}
\label{Eq:Critical_Values}
\tilde{r}_{+c} = 0.435773, \quad \tilde{q}_c=0.142336, \quad \text{and} \quad \tilde{T}_c=0.264695
\end{equation}

\begin{figure}[h!]
	\centerline{
	\includegraphics[scale=0.5]{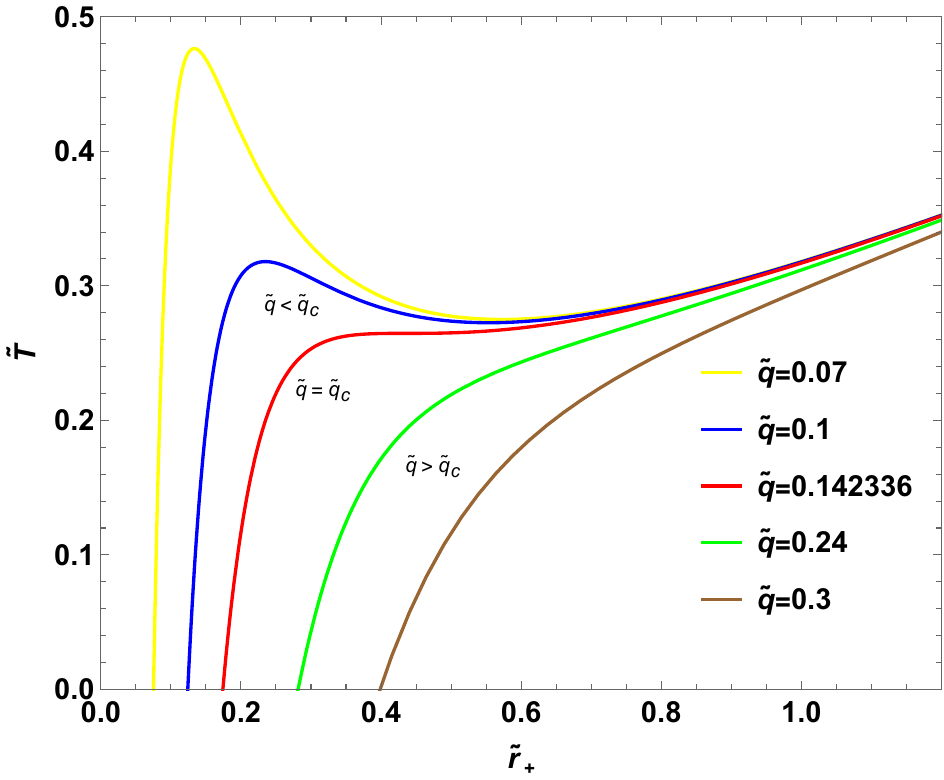}}
	\caption	{Hawking temperature as a function of horizon radius for different values of $\tilde{q}$ above (green and brown) and below (yellow and blue) the critical point $\tilde{q}_c=0.142336$ (red).}	
	\label{Tem_vs_r}
	\end{figure}

 Using the Hawking temperature expression \eqref{Eq:HawTem} and the scaling \eqref{Eq:AdsScaling}, we plot temperature as a function of horizon radius for different values of $\tilde{q}$ as shown in \autoref{Tem_vs_r}. From this figure, we find different black hole solutions with different horizon radii $\tilde{r}_+$ below the critical point $\tilde{q}_c$. Above $\tilde{q}_c$, there exists a single black hole solution for a range of horizon radius and temperature.

\begin{figure}[ht]
	\centering
		\begin{subfigure}{0.55\textwidth}
			\centering
			\includegraphics[width=0.9\linewidth]{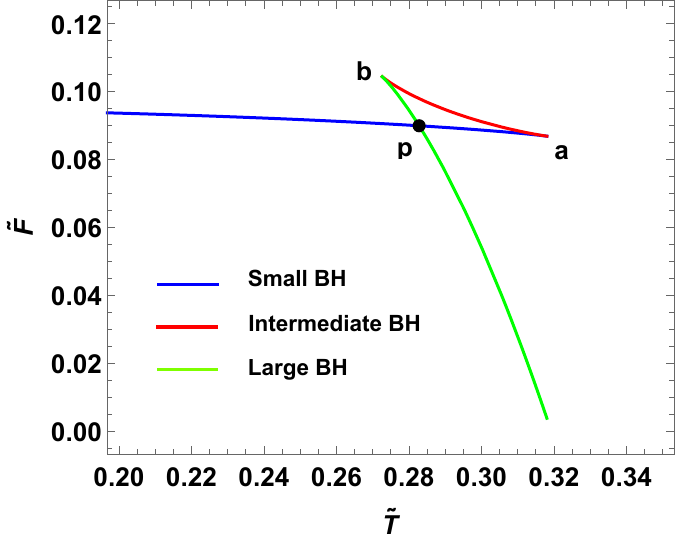}
			\caption{For $\tilde{q}=0.1<\tilde{q}_c$}
			\label{}
		\end{subfigure}%
		\begin{subfigure}{0.55\textwidth}
			\centering
			\includegraphics[width=0.9\linewidth]{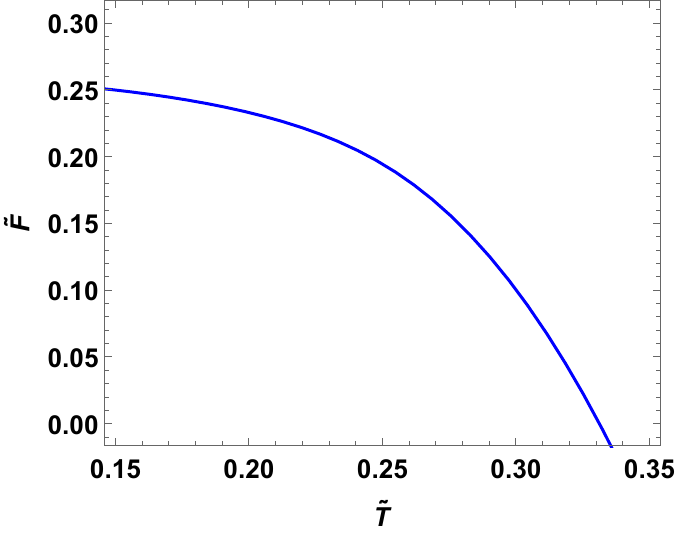}
			\caption{For $\tilde{q}=0.25>\tilde{q}_c$}
			\label{}
		\end{subfigure} \\
	\caption{Gibbs free energy as a function of $\tilde{T}$ for fixed values of $\tilde{q}$.}
	\label{Gibbs_Free_Energy}
	\end{figure}

Now, to study the phase transition, we use Gibbs free energy given by \eqref{Eq:Gibb_Free_Energy}. We express the horizon radius $\tilde{r}_+$ as a function of Hawking temperature $\tilde{T}$ using \eqref{Eq:HawTem} and find that $\tilde{r}_+(\tilde{T})$ is multivalued. Then, we put $\tilde{r}_+(\tilde{T})$ in \eqref{Eq:Gibb_Free_Energy} and finally obtain the rescaled free energy $\tilde{F}$ as a function of $\tilde{T}$ and $\tilde{q}$. The Gibbs free energy thus obtained are shown in \autoref{Gibbs_Free_Energy} with fixed $\tilde{q}$. When $\tilde{q}$ is smaller than the $\tilde{q}_c$ we have three black hole solutions namely, small BH, intermediate BH and large BH. These three black hole solutions can coexist for $\tilde{T}_b<\tilde{T}<\tilde{T}_a$, where $\tilde{T}_b$ and $\tilde{T}_a$ are the temperatures at the point $b$ and $a$ respectively. The temperature at the point $p$ represents the phase transition point $(\tilde{T}_p=0.282789)$. When $\tilde{q}$ is greater than $\tilde{q}_c$, there will be no phase transition as we have only a single black hole solution. 

\section{Lyapunov exponent and phase transition}
\label{LyapunovExponent}
The concept of Lyapunov exponent is widely used in the field of dynamical systems and chaos theory. It quantifies how quickly nearby trajectories in a system either move apart (diverge) or come together (converge) over time. In this section, we intend to study the Lyapunov exponents of massless and massive particles in an unstable circular orbit on the equatorial plane around the Hayward AdS black hole. 
While the computation of the Lyapunov exponent is well known, we will provide a brief overview for the convenience of the readers.
 To begin, we commence with the Lagrangian, focusing on the case where $\theta=\pi/2$. This can be expressed as:
\begin{equation}
\label{Eq:Lagrangian}
\mathcal{L}=\frac{1}{2} \Bigl\{-f(r)\Big(\frac{d t}{d \tau}\Big)^2 + \frac{1}{f(r)}\Big(\frac{d r}{d \tau}\Big)^2+r^2 \Big(\frac{d \phi}{d \tau}\Big)^2
\Bigr\},
\end{equation}
where $\tau$ is the proper time and $f(r)$ is given by \eqref{Eq:Fr}.
From the Lagrangian the canonical momenta of the particle can be easily worked out as:
\begin{equation}
\label{Eq:Canonical_monemta}
p_t=\frac{\partial \mathcal{L}}{\partial \dot{t}}= - f(r) \dot{t} = - E ,\quad
p_r=\frac{\partial \mathcal{L}}{\partial \dot{r}}= \frac{1}{f(r)} \dot{r} , \quad
p_\phi=\frac{\partial \mathcal{L}}{\partial \dot{\phi}}= r^2 \dot{\phi}=L,
\end{equation}
where $E$ and $L$ are the energy and angular momentum of the particle. Also, the dots represent the derivatives with respect to proper time $\tau$. From \eqref{Eq:Canonical_monemta} we can find that 
\begin{equation}
\label{Eq:Transformation}
\dot{t}=\frac{E}{f(r)} \quad \text{and} \quad \dot{\phi} =\frac{L}{r^2}.
\end{equation}
Then, we calculate the Hamiltonian as 
\begin{equation}
\label{Eq:Hamiltonian}
\begin{aligned}
2\mathcal{H} &= 2(p_t \dot{t} + p_r \dot{r} + p_\phi \dot{\phi}-\mathcal{L}) \\
&= -f(r)\dot{t}^2 +\frac{\dot{r}^2}{f(r)}+r^2\dot{\phi}^2 \\
&= -\frac{E^2}{f(r)}+\frac{\dot{r}^2}{f(r)}+\frac{L^2}{r^2} = -\delta_1,
\end{aligned}
\end{equation}
where we have used \eqref{Eq:Transformation}. For timelike geodesic $\delta_1=1$ and for null geodesic $\delta_1=0$. 
Using the definition of effective potential for radial motion, $V_r=-\dot{r}^2$ in \eqref{Eq:Hamiltonian}, we find that 
\begin{equation}
\label{Eq:Effective_Potential}
V_r= f(r)\Big[ \delta_1 -\frac{E^2}{f(r)} +\frac{L^2}{r^2} \Bigr]
\end{equation} 
Now, if we write the angular momentum $L$ in terms of the effective potential $V_r$ (setting $E=0$) and plug it into \eqref{Eq:Hamiltonian} then, the Hamiltonian can be expressed as
\begin{equation}
\begin{aligned}
\mathcal{H}&=\frac{V_r-E^2}{2 f(r)}+\frac{\dot{r}^2}{f(r)}-\frac{\delta_1}{2} \\
&=\frac{V_r-E^2}{2 f(r)}+\frac{f(r)p_r^2}{2}-\frac{\delta_1}{2} 
\end{aligned}
\end{equation}
The equations of motion in proper time configuration can be derived from the Hamiltonian as:
\begin{equation}
\begin{aligned}
& \dot{r}=\frac{\partial \mathcal{H}}{\partial p_r}=f(r)p_r  , \\
& \dot{p_r}=-\frac{\partial \mathcal{H}}{\partial r}= \frac{V_r-E^2 }{2 f(r)^2} f'(r)-\frac{1}{2} p^2 f'(r)-\frac{V'_r}{2 f(r)},
\end{aligned} 
\end{equation}
where the primes denote the derivatives with respect to $r$. Now, we can linearize these equations of motion about the circular orbit $r_0$ and calculate the linear stability matrix $K$ in terms of the coordinate time $t$ as:
\begin{equation}
\label{Eq:Stability_matrix}
 \begin{pmatrix}
0 & \frac{f(r_0)}{\dot{t}} \\
-\frac{V_r''(r_0)}{2f(r_0)\dot{t}} & 0 
\end{pmatrix}  
\end{equation}
The eigenvalue of \eqref{Eq:Stability_matrix} gives the Lyapunov exponent
\begin{equation}
\label{Eq:General_Lyapunov}
\lambda=\sqrt{-\frac{V''_r(r_0)}{2 \dot{t}^2}},
\end{equation}
where we have dropped $\pm$ for simplicity. Also, $\lambda$ is real when $V''_r(r_0)<0$.
\subsection{Massless particle (null geodesic)}
\begin{figure}[ht!]
	\centerline{
	\includegraphics[scale=0.6]{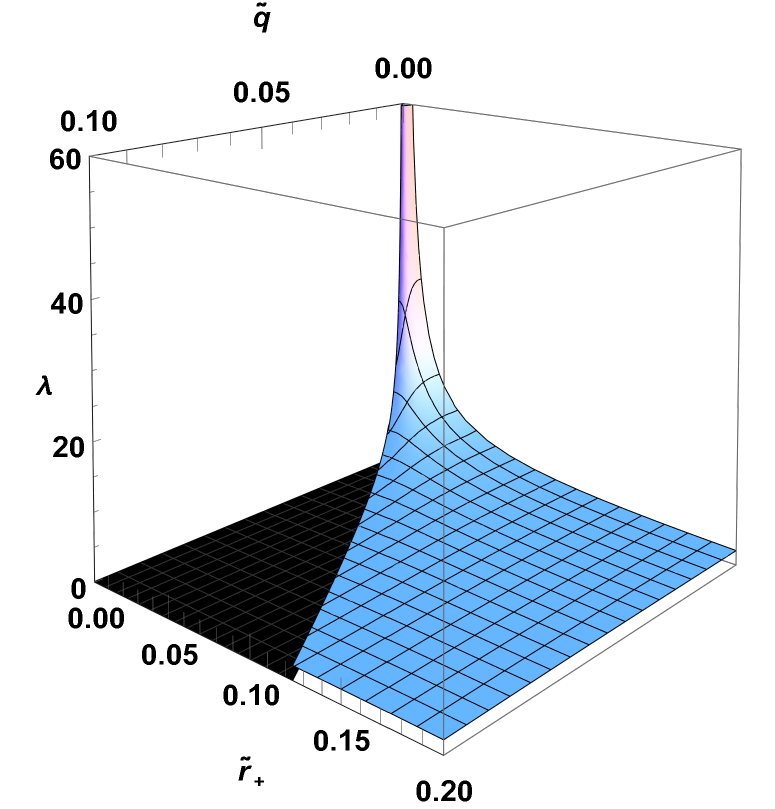}}
	\caption	{Lyapunov exponent $\lambda$ as a function of $\tilde{q}$ and $\tilde{r}_+$ for unstable circular null geodesics. Black region represents non-physical region.}	
	\label{Null_Geo_Lyapunov_3d}
	\end{figure}

To calculate the Lyapunov exponent for massless particle we set $\delta_1=0$. The condition for unstable geodesic is $V_r(r_0)=V'_r(r_0)=0$ and $V''_r(r_0)<0$. From \eqref{Eq:Effective_Potential} and using these conditions we can find that 
\begin{equation}
\frac{E}{L}=\frac{\sqrt{f(r_0)}}{r_0}
\end{equation}
Plugging this in \eqref{Eq:Transformation} we get
\begin{equation}
\label{Eq:Nul_t_dot}
\dot{t}=\frac{L}{r_0\sqrt{f(r_0)}}
\end{equation}
Now, we find the radius of the unstable circular orbit by the condition $V'_r(r_0)=0$ and $V''_r(r_0)<0$. To work this out, we calculate the first derivative of $V_r(r_0)$ and equate it to zero with $\delta_1=0$. i.e.,
\begin{equation}
V'_r(r_0)=-\frac{2 L^2 \Bigl\{r_0^5 \left(r_0-3 \tilde{M}\right)+2 r_0^3 \tilde{q}^3+\tilde{q}^6\Bigr\}}{r_0^3 \left(\tilde{q}^3+r_0^3\right){}^2}=0
\end{equation} 
Then, we solve the obtained equation for $r_0$ and check for $V''_r(r_0)<0$. 
Finally, we use the expression for mass \eqref{Eq:Mass} with scaling \eqref{Eq:AdsScaling} to find $r_0$ as a function of $\tilde{q}$ and $\tilde{r}_+$. We observe that $r_0$ is independent of the angular momentum $L$. The explicit expression is not shown intentionally for simplicity. Also, the second derivative of the effective potential is given as 
\begin{equation}
\label{Eq:Second_Derivative_of_Potential}
V''_r(r_0)=\frac{6 L^2 \Bigl\{ r_0^5 \tilde{q}^3 \left(2 \tilde{M}+3 r_0\right)+r_0^8 \left(r_0-4 \tilde{M}\right)+3 r_0^3 \tilde{q}^6+\tilde{q}^9\Bigr\} }{r_0^4 \left(\tilde{q}^3+r_0^3\right){}^3}
\end{equation}
The Lyapunov exponent $\lambda$ is then calculated for $r_0$ using \eqref{Eq:Nul_t_dot}, \eqref{Eq:Second_Derivative_of_Potential} in \eqref{Eq:General_Lyapunov}. We observe that $\lambda$ depends on $\tilde{q}$ and $\tilde{r}_+$ and it is independent of angular momentum $L$. The plot of Lyapunov exponent $\lambda$ for null geodesic as a function of $\tilde{q}$ and $\tilde{r}_+$ is shown in \autoref{Null_Geo_Lyapunov_3d}. In the figure, no black hole solution exists in the black region. This can be understood if we observe the Hawking temperature $\tilde{T}$ which is negative for $\sqrt[3]{2} \tilde{q}<\sqrt[3]{3 \tilde{r}^5+\tilde{r}^3}$.

\begin{figure}[h!]
	\centerline{
	\includegraphics[scale=0.6]{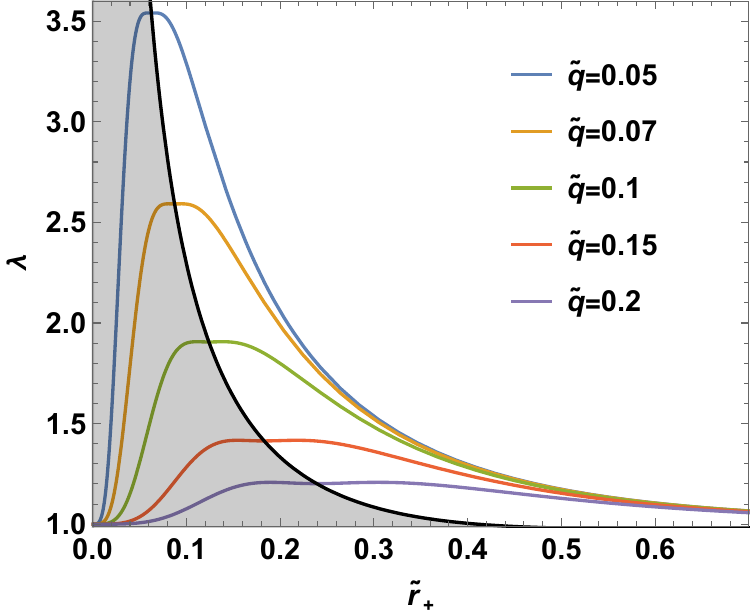}}
	\caption	{2D view of Lyapunov exponent curves for unstable circular null geodesics with different values of $\tilde{q}$. Gray area represents the non-physical region.}	
	\label{Null_Geo_Lyapunov_2d}
	\end{figure}

For a better understanding of the Lyapunov exponent for unstable circular null geodesics, we plot $\lambda$ in a 2D plane for fixed values of $\tilde{q}$ in \autoref{Null_Geo_Lyapunov_2d}. The gray area in the figure represents a non physical region because of the Hawking temperature being negative $(\tilde{T}<0)$. On the black curve, the temperature is zero $(\tilde{T}=0)$. The figure also shows that for smaller value of $\tilde{r}$, the Lyapunov exponent $\lambda$ increases as we decrease the values of $\tilde{q}$. As we gradually increase $\tilde{r}_+$, the Lyapunov exponent curves for different values of $\tilde{q}$, start to coalesce. In fact, as $\tilde{r}_+$ or $\tilde{q}$ tends towards infinity, $\lambda$ approaches to $1$.

\begin{figure}[ht]
	\centering
		\begin{subfigure}{0.53\textwidth}
			\centering
			\includegraphics[width=0.9\linewidth]{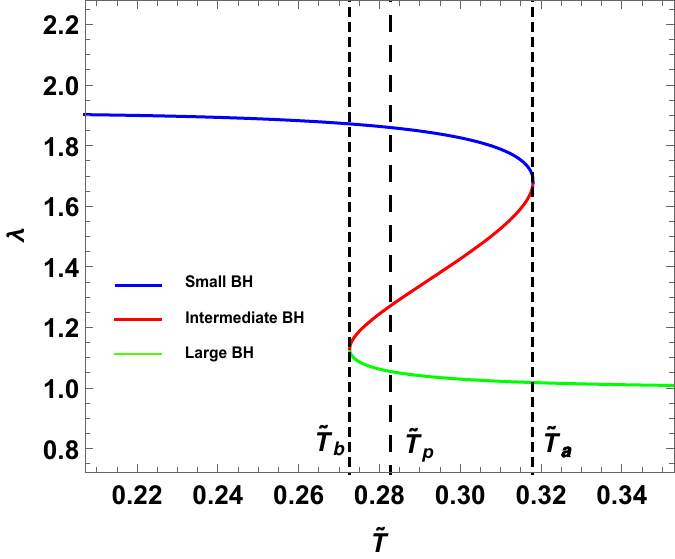}
			\caption{For $\tilde{q}=0.1<\tilde{q}_c$}
			\label{Lyapunov_vs_T_Below_Critical}
		\end{subfigure}%
		\begin{subfigure}{0.55\textwidth}
			\centering
			\includegraphics[width=0.9\linewidth]{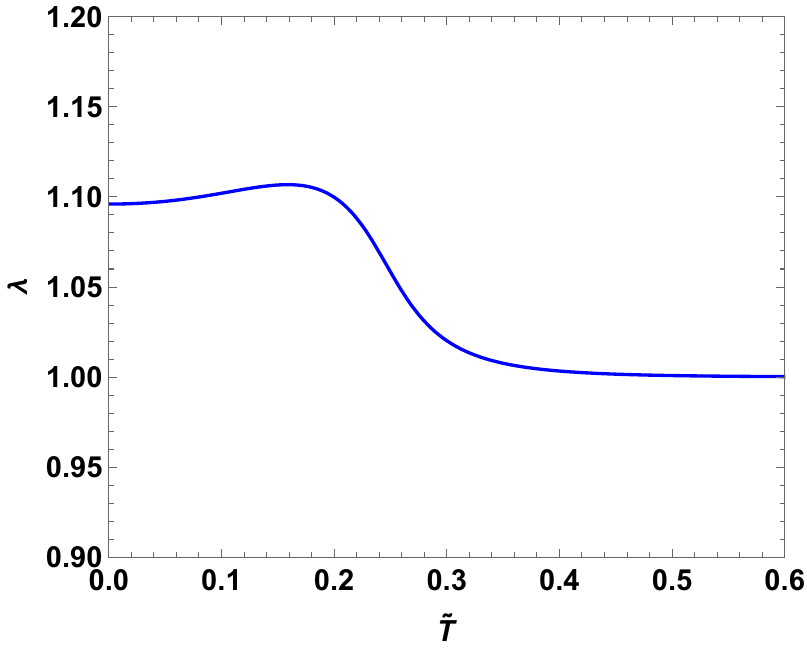}
			\caption{For $\tilde{q}=0.25>\tilde{q}_c$}
			\label{Lyapunov_vs_T_Above_Critical}
		\end{subfigure} \\
	\caption{Lyapunov exponent as a function of temperature $\tilde{T}$ for unstable circular null geodesics.}
	\label{Lyapunov_vs_T}
	\end{figure}

We can write the Lyapunov exponent $\lambda$ in terms of temperature $\tilde{T}$ by expressing $\tilde{r}_+$ as a function of $\tilde{T}$ from the Hawking temperature expression. As discussed earlier $\tilde{r}_+(\tilde{T})$ is multivalued and hence we obtain multiple functions of Lyapunov exponent which defines different phases of the Hayward AdS black hole. We have shown $\lambda$ as a function of $
\tilde{T}$ for fixed values of $\tilde{q}$ in \autoref{Lyapunov_vs_T}. The left figure (\autoref{Lyapunov_vs_T_Below_Critical}) shows $\lambda$ for $\tilde{q}<\tilde{q}_c=0.142336$. Here, we have three different regions of Hayward AdS black hole which are small black hole (blue), intermediate black hole (red) and large black hole (green). The point $\tilde{T}_p=0.282789$ is the phase transition temperature. For $\tilde{T}_b<\tilde{T}<\tilde{T}_a$ the Lyapunov exponent $\lambda$ has three branches and all the three black hole solutions (small BH, intermediate BH and large BH) coexist in this region. As temperature is raised from $\tilde{T}_b$ to $\tilde{T}_a$, $\lambda$ for small black hole and large black hole decreases slightly and $\lambda$ for intermediate black hole increases from their respective positions. Also, $\lambda \to 1 $ when $\tilde{T} \to \infty$. For a $\tilde{q}$ value greater than $\tilde{q}_c$, say $\tilde{q}=0.25$, the Lyapunov exponent $\lambda$ is single valued for any value of temperature $\tilde{T}$ as shown in \autoref{Lyapunov_vs_T_Above_Critical}. In this case, there exists a single black hole solution and phase transition is not possible. From \autoref{Lyapunov_vs_T_Above_Critical}, we observe that the trend of $\lambda$ initially exhibits a slight increase, followed by a decrease as we gradually raise $\tilde{T}$. As $\tilde{T}$ approaches infinity, $\lambda$ tends toward $1$. 

\begin{figure}[h!]
	\centerline{
	\includegraphics[scale=0.65]{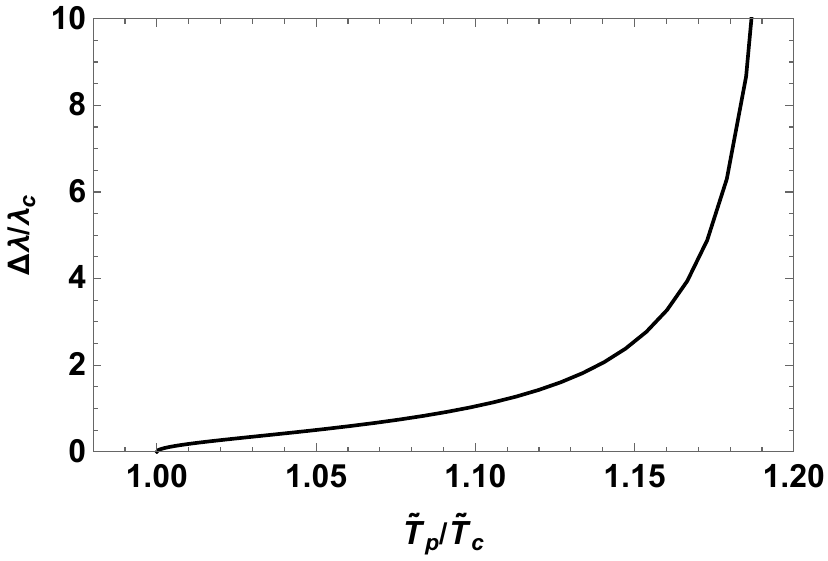}}
	\caption	{$\Delta \lambda  / \lambda_c$ vs $\tilde{T}_p / \tilde{T}_c$ curve for null geodesics.}	
	\label{DelLambda_Massless_FullRange}
	\end{figure}

Now, we study the difference of Lyapunov exponents in the phase transition point of Hayward AdS black hole. At the small- large phase transition point $p$, the Lyapunov exponent for small and large black hole is respectively denoted as $\lambda_s$ and $\lambda_l$. With different values of $\tilde{q}$, the phase transition temperature $\tilde{T}_p$ changes and for these values of $\tilde{T}_p$ we calculate the difference of Lyapunov exponents $\Delta \lambda=\lambda_s-\lambda_l$. The phase transition vanishes at the critical point $\tilde{q}=\tilde{q}_c$ as the two extreme points of $\tilde{T}$ vs $\tilde{r}_+$ curve coincides. At this point $\tilde{T}_p=\tilde{T}_c$ and $\lambda_s=\lambda_l=\lambda_c$ which results $\Delta\lambda=0$. The critical value of Lyapunov exponent $\lambda_c$ can be calculated by inserting the critical values given in \eqref{Eq:Critical_Values} and it is found as $\lambda_c=1.2118$. We represent the $\Delta \lambda  / \lambda_c$ vs $\tilde{T}_p / \tilde{T}_c$ curve in \autoref{DelLambda_Massless_FullRange}.
In this figure, we see that the change in Lyapunov exponent $\Delta\lambda$ is non-zero at the phase transition. As the phase transition temperature $\tilde{T}_p$ slowly moves towards the critical temperature $\tilde{T}_c$, the difference in Lyapunov exponent $\Delta\lambda$ non-linearly decreases.  At the critical point $\tilde{T}_p / \tilde{T}_c=1$ and $\Delta \lambda  / \lambda_c=0$. Such behaviors of the parameter $\Delta\lambda$ indicate that $\Delta\lambda$ acts as an order parameter.

\begin{figure}[h!]
	\centerline{
	\includegraphics[scale=0.65]{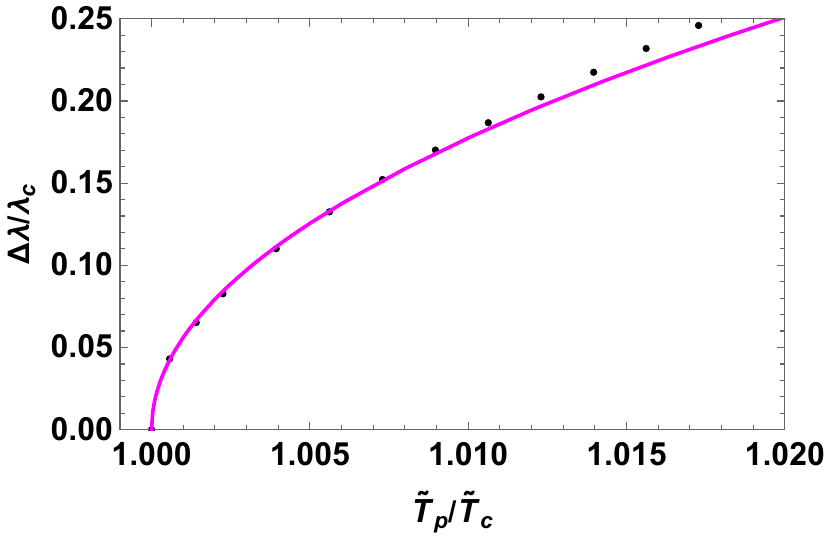}}
	\caption	{$\Delta \lambda  / \lambda_c$ vs $\tilde{T}_p / \tilde{T}_c$ curve for null geodesics near critical point. The black dots represents $\Delta\lambda$ (scaled with $\lambda_c$) for different values of $\tilde{q}$. The magenta curve is for $\Delta \lambda/\lambda_c=1.77499\sqrt{\tilde{T}_p / \tilde{T}_c-1}$}	
	\label{DelLambda_Massless}
	\end{figure}

To study the critical behavior of $\Delta\lambda$ we calculate the critical exponent, a numerical value that characterizes the behavior of a physical system near its critical point. The relation of critical exponent $\delta$ and $\Delta \lambda$ is defined as \cite{Lyu:2023sih}:
\begin{equation}
\Delta\lambda\equiv \lambda_s-\lambda_l\sim|\tilde{T}-\tilde{T}_c|^\delta.
\end{equation}
To calculate $\delta$ we follow the method provided in \cite{Banerjee:2012zm}. We rewrite the horizon radius at phase transition point and the Hawking temperature as
\begin{equation}
\label{Eq:rpExpand}
\tilde{r}_p=\tilde{r}_c(1+\Delta),
\end{equation}
and 
\begin{equation}
\label{Eq:HTExpand}
\tilde{T}(\tilde{r}_+)=\tilde{T}_c(1+\epsilon),
\end{equation}
where $|\Delta|\ll1$ and $|\epsilon|\ll1$.
The Lyapunov exponents can be expanded using Taylor series about the critical point $\tilde{r}_c$ as 
\begin{equation}
\label{Eq:LambdaTaylor}
\lambda=\lambda_c+\left[\frac{\partial\lambda}{\partial \tilde{r}_+}\right]_c d\tilde{r}_+ + \mathcal{O}(\tilde{r}_+),
\end{equation}
where the subscript $``c"$ is used to represent values at the critical point.
Using \eqref{Eq:rpExpand} and \eqref{Eq:LambdaTaylor} we can find
\begin{equation}
\label{Eq:DelLambdaTaylor}
\Delta\tilde{\lambda}=\frac{\Delta\lambda}{\lambda_c}=\frac{\lambda_s-\lambda_l}{\lambda_c}=\frac{\tilde{r}_c}{\lambda_c}\left[ \frac{\partial \lambda}{\partial \tilde{r}_+}  \right]_c (\Delta_s-\Delta_l),
\end{equation}
where the subscript $``s"$ and $``l"$ represents small and large black hole branch. Here, we have also used $\lambda_s(\tilde{r}_c)-\lambda_l(\tilde{r}_c)=0$. Similarly, we can Taylor expand Hawking temperature about the critical point $\tilde{r}_c$ and find
\begin{equation}
\label{Eq:HTTaylor}
\tilde{T}=\tilde{T}_c+\frac{\tilde{r}_c^2}{2}\left[  \frac{\partial^2\tilde{T}}{\partial \tilde{r}_+^2} \right]_c \Delta^2,
\end{equation}
where we have omitted the higher order terms. Finally, using \eqref{Eq:DelLambdaTaylor} and \eqref{Eq:HTTaylor} we can obtain 
\begin{equation}
\label{Eq:DeltaCriticalExponent}
\Delta\tilde{\lambda}=k\sqrt{t-1},
\end{equation}
where $t=\frac{\tilde{T}}{\tilde{T}_c}$ and
\begin{equation}
k=\frac{\sqrt{\tilde{T}_c}}{\lambda_c}\left[  \frac{1}{2} \frac{\partial^2\tilde{T}}{\partial \tilde{r}_+^2} \right]^{-1/2}_c \left[ \frac{\partial\Delta\lambda}{\partial \tilde{r}_+}  \right]_c.
\end{equation}
Therefore, the critical exponent $\delta$ of $\Delta\lambda$ near the critical point is $1/2$ which is same as that of the order parameter in VdW fluid.  In \autoref{DelLambda_Massless} we are focusing on the parameter $\Delta \lambda$  near the critical point $\tilde{T}_c$. The black dot represents the parameter $\Delta\lambda$ (scaled with $\lambda_c$) for massless particles. We have calculated the value of $k$ numerically and subsequently, using \eqref{Eq:DeltaCriticalExponent} find that
\begin{equation}
\label{Eq:DeltaFittingEq}
\Delta\tilde{\lambda}=\frac{\Delta \lambda}{\lambda_c}=1.77499\sqrt{t-1}=1.77499\sqrt{\tilde{T}_p / \tilde{T}_c-1},
\end{equation}
which is represented by the magenta curve in \autoref{DelLambda_Massless} and this serves a good fit for $\Delta \lambda  / \lambda_c$ (black dots in the figure). This further confirms that the critical exponent for $\Delta\lambda$ is $1/2$.

\subsection{Massive particle (timelike geodesic)}
\begin{figure}[h!] 
	\centerline{
	\includegraphics[scale=0.65]{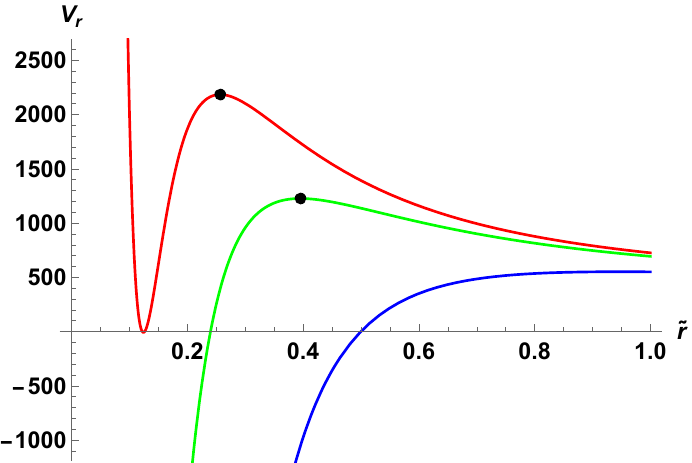}}
	\caption	{Effective potential $V_r$ as a function $\tilde{r}$ for $\tilde{q}=0.1$. Red, green and blue curves are respectively for $\tilde{r}_+=0.122$, $\tilde{r}_+=0.24$ and $\tilde{r}_+=0.5$. }	
	\label{Effective_Potential}
	\end{figure}

For timelike geodesics we chose $\delta_1=1$. The condition $V(r_0)=V'(r_0)=0$, provides us the following relations for energy and angular momentum as
\begin{equation}
E^2=\frac{2 f\left(r_0\right){}^2}{2 f\left(r_0\right)-r_0 f'\left(r_0\right)},
\end{equation}
and
\begin{equation}
L^2=\frac{r_0^3 f'\left(r_0\right)}{2 f\left(r_0\right)-r_0 f'\left(r_0\right)}.
\end{equation}
Therefore, from \eqref{Eq:Transformation} we obtain
\begin{equation}
\label{Eq:Timelike_t_dot}
\dot{t}=\frac{1}{\sqrt{f\left(r_0\right)-\frac{1}{2} r_0 f'\left(r_0\right)}}
\end{equation}
The radius $r_0$ of the unstable circular geodesic is calculated from $V'(r_0)=0$ with $\delta_1=1$ and checked for $V''(r_0)<0$. Unlike the null geodesics case, here $r_0$ depends on the angular momentum $L$. We have not shown the explicit form of $r_0$ for simplicity. Also, the second derivative of potential is 
\begin{equation}
\label{Eq:Second_Derivative_of_Potential_Timelike}
V''_r(r_0)=\frac{
\begin{aligned}
6 \tilde{L}^2 \Bigl\{ \tilde{q}^3 \tilde{r}^5 \left(2 \tilde{M}+3 \tilde{r}\right)+\tilde{r}^8 \left(\tilde{r}-4 \tilde{M}\right)+3 \tilde{q}^6 \tilde{r}^3+\tilde{q}^9\Bigr\} \\
+2 \tilde{r}^4 \Bigl\{ \left(\tilde{q}^3+\tilde{r}^3\right)^3-2 \tilde{M} \left(-7 \tilde{q}^3 \tilde{r}^3+\tilde{q}^6+\tilde{r}^6\right)\Bigr\} 
\end{aligned}
}{\tilde{r}^4 \left(\tilde{q}^3+\tilde{r}^3\right)^3}.
\end{equation}
The effective potential $V_r$ for unstable timelike null geodesics can be written as a function of $\tilde{r}$, $\tilde{r}_+$ and $\tilde{q}$ using the expression of $f(r)$ and $\tilde{M}$ in \eqref{Eq:Effective_Potential}. The $V_r-\tilde{r}$ relation is shown in \autoref{Effective_Potential} for different values of $\tilde{r}_+$ with $\tilde{q}=0.1$. Here, we have set $L=20l$ and $E=0$. In the figure, the black dots represent the maximum of the effective potential for which $V''_r<0$ corresponding to unstable equilibria. The minimum of $V_r$ for which $V''_r>0$ corresponds to stable equilibria. The figure also shows that the maximum of $V_r$ decreases with the increase of $\tilde{r}_+$ and for $\tilde{r}_+=0.5$ there is no maximum. This implies that the unstable timelike geodesic will disappear for large value of $\tilde{r}_+$.

\begin{figure}[ht!]
	\centerline{
	\includegraphics[scale=0.75]{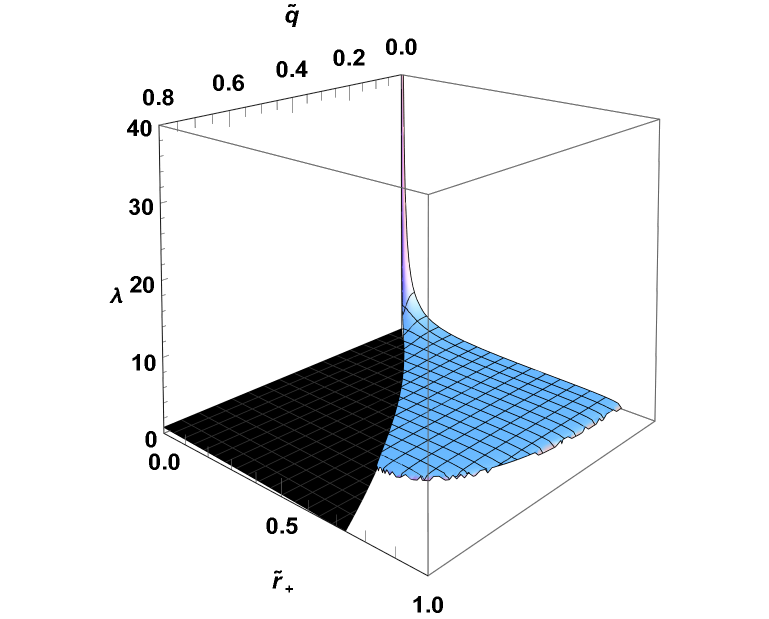}}
	\caption	{Lyapunov exponent $\lambda$ as a function of $\tilde{q}$ and $\tilde{r}_+$ for unstable circular timelike geodesics. Black region represents non-physical region.}	
	\label{Timelike_Geo_Lyapunov_3d}
	\end{figure}

\begin{figure}[h!]
	\centerline{
	\includegraphics[scale=0.65]{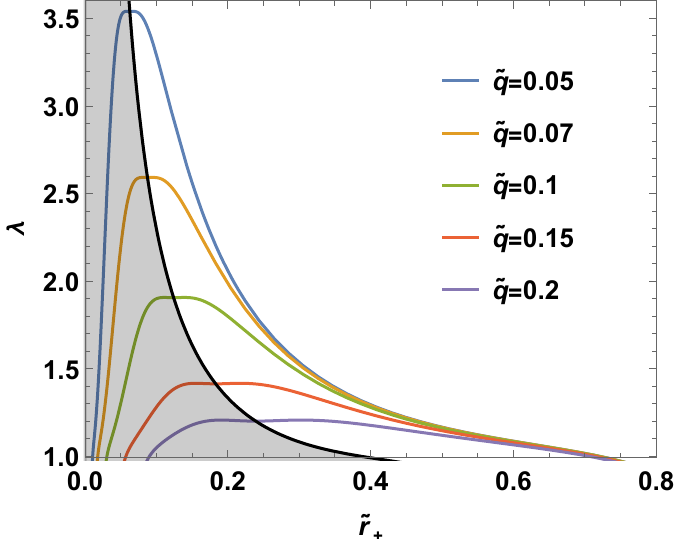}}
	\caption	{2D view of Lyapunov exponent curves for unstable circular geodesics with different values of $\tilde{q}$. Gray area represents the non-physical region.}	
	\label{Timelike_Geo_Lyapunov_2d}
	\end{figure}

Using \eqref{Eq:Timelike_t_dot} and \eqref{Eq:Second_Derivative_of_Potential_Timelike} in \eqref{Eq:General_Lyapunov} we can calculate the Lyapunov exponent $\lambda$ for unstable circular timelike geodesics. In this case, $\lambda$ depends on $L$, $\tilde{q}$ and $\tilde{r}_+$.  For simplicity we have avoided writing the explicit form of  $\lambda$. The three dimensional representation of the Lyapunov exponent $\lambda$ for unstable circular timelike geodesic is shown in \autoref{Timelike_Geo_Lyapunov_3d} where we have chosen $L=20l$. In the figure, the black area represents non-physical region with negative temperature. The unstable region exist for smaller values of $\tilde{r}_+$ and $\lambda$ in this region is represented by the blue surface. In the white region, there is no unstable timelike circular orbits and therefore, $\lambda$ vanishes in this region.
The two dimensional view of the Lyapunov exponent for timelike geodesic is shown in the \autoref{Timelike_Geo_Lyapunov_2d}. Here, the non-physical region with negative temperature is shaded as gray.

\begin{figure}[ht]
	\centering
		\begin{subfigure}{0.53\textwidth}
			\centering
			\includegraphics[width=0.9\linewidth]{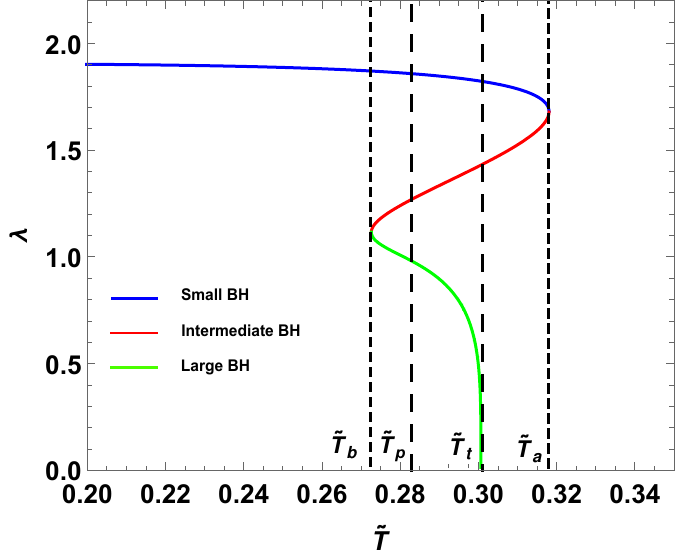}
			\caption{For $\tilde{q}=0.1<\tilde{q}_c$}
			\label{Lyapunov_vs_T_Timelike_Below_Critical}
		\end{subfigure}%
		\begin{subfigure}{0.55\textwidth}
			\centering
			\includegraphics[width=0.85\linewidth]{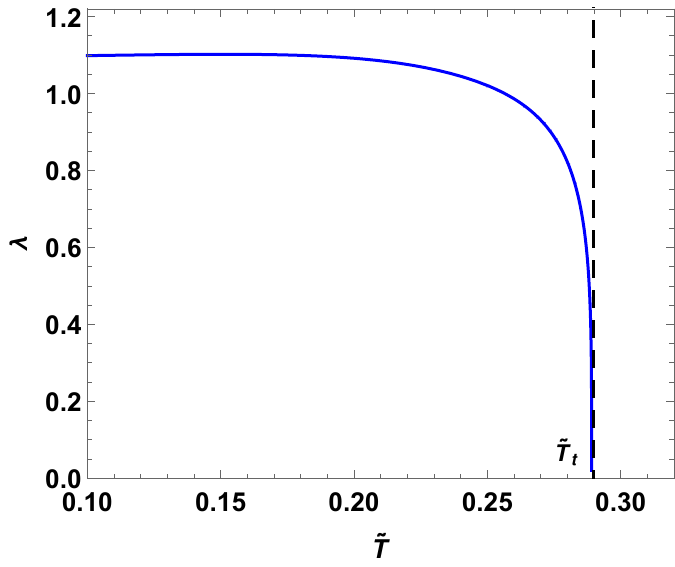}
			\caption{For $\tilde{q}=0.25>\tilde{q}_c$}
			\label{Lyapunov_vs_T_Timelike_Above_Critical}
		\end{subfigure} \\
	\caption{Lyapunov exponent as a function of temperature $\tilde{T}$ for unstable circular timelike geodesics.}
	\label{Lyapunov_vs_T_Timelike}
	\end{figure}

To study the relationship between phase transition and Lyapunov exponent, we write $\lambda$ in terms of $\tilde{T}$ using the Hawking temperature.  The Lyapunov exponent $\lambda$ for $L=20l$ is shown in \autoref{Lyapunov_vs_T_Timelike}. The left figure (\autoref{Lyapunov_vs_T_Timelike_Below_Critical}) shows $\lambda$ for $\tilde{q}=0.1$ which is below the critical value $\tilde{q}_c=0.142336$. For this value of $\tilde{q}$, the Lyapunov exponent $\lambda$ is multivalued and it has three branches. These branches corresponds to three different phases or three black hole solutions of Hayward AdS black hole which can coexist for $\tilde{T}_b<\tilde{T}<\tilde{T}_t$ .
The phase transition from small black hole to large black hole occurs at the temperature $\tilde{T}=\tilde{T}_p=0.2871785$. For $\tilde{q}=0.25>\tilde{q}_c$ there is no phase transition and the Lyapunov exponent $\lambda$ is found to be single valued. Unlike the massless particles case, $\lambda$ for massive particles vanishes and become zero at the temperature point $\tilde{T}=\tilde{T}_t$.

\begin{figure}[ht]
	\centering
		\begin{subfigure}{0.52\textwidth}
			\centering
			\includegraphics[width=0.9\linewidth]{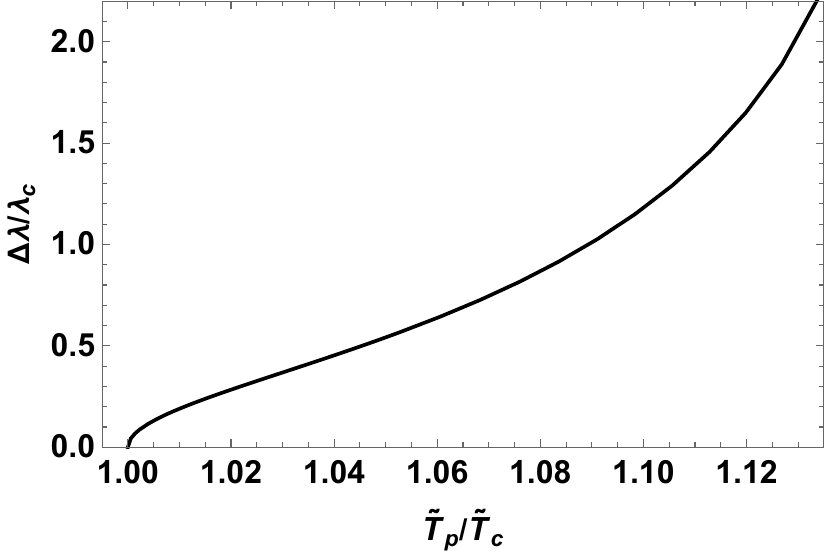}
			\caption{Large range view of $\Delta \lambda  / \lambda_c$ vs $\tilde{T}_p / \tilde{T}_c$ curve.}
			\label{DelLambda_Massive_FullRange}
		\end{subfigure}%
		\begin{subfigure}{0.55\textwidth}
			\centering
			\includegraphics[width=0.9\linewidth]{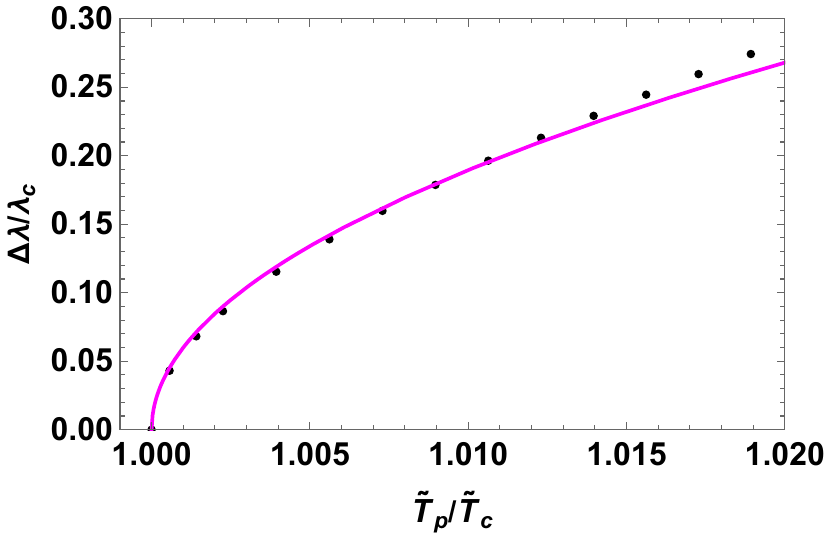}
			\caption{$\Delta \lambda  / \lambda_c$ vs $\tilde{T}_p / \tilde{T}_c$ curve near the critical point.}
			\label{DelLambda_Massive}
		\end{subfigure} \\
	\caption{$\Delta \lambda  / \lambda_c$ vs $\tilde{T}_p / \tilde{T}_c$ curve for timelike geodesics. The black dots represents $\Delta\lambda$ (scaled with $\lambda_c$) for different values of $\tilde{q}$. The magenta curve is for $\Delta \lambda/\lambda_c=1.89539\sqrt{\tilde{T}_p / \tilde{T}_c-1}$.}
	\label{DelLambda_Massive_With_FullRange}
	\end{figure}

Now, we calculate the change in the difference of Lyapunov exponent $\Delta\lambda$ for different values of $\tilde{q}$. The plot of  $\Delta \lambda  / \lambda_c$ vs  $\tilde{T}_p / \tilde{T}_c$ is shown in \autoref{DelLambda_Massive_FullRange}. The figure shows that the parameter $\Delta\lambda$ is non-zero at the phase transition and it non-linearly decrease as the phase transition temperature $\tilde{T}_p$ slowly approaches to the critical temperature $\tilde{T}_c$. At the critical point $\tilde{T}_p / \tilde{T}_c=1$ and $\Delta \lambda  / \lambda_c=0$.  The critical behavior of $\Delta\lambda$ is shown in \autoref{DelLambda_Massive}. Near the critical point, $\Delta\lambda/\lambda_c$ (black dots in the figure) is well represented by
\begin{equation}
\frac{\Delta \lambda}{\lambda_c}=1.89539\sqrt{\tilde{T}_p / \tilde{T}_c-1},
\end{equation}
which confirms that the critical exponent of $\Delta\lambda$ is $1/2$.

\newpage
\newpage
\section{Conclusion}
\label{Conclusion}
In this paper, we have studied the link between Lyapunov exponents and the phase structure of 4D Hayward AdS black hole. We have calculated the Lyapunov exponents for massless and massive particles in an unstable circular orbit of the black hole in the equatorial plane and study its behavior. 

For the massless particles, below the critical value of $\tilde{q}_c$, the Lyapunov exponent $\lambda$ has three different branches each of them corresponding to three different phases (SBH, IBH and LBH) of the Hayward AdS black hole. Above the critical value of $\tilde{q}_c$, the Lyapunov exponent $\lambda$ has a single branch. In this case, there is no phase transition. Which implies that $\lambda$ is multivalued when there is a phase transition. Also, $\lambda$ tends to $1$ when temperature $\tilde{T}$ tends to infinity implying that there is no terminating temperature for $\lambda$ in case of unstable circular null geodesics in Hayward AdS black hole.

The motion of massive particles around the Hayward AdS black hole in unstable circular orbit is defined by the timelike geodesics. We observe that below the critical value of $\tilde{q}_c$, the Lyapunov exponent $\lambda$ is multivalued and its three different branches corresponds to three different phases of Hayward AdS black hole. Above the critical value of $\tilde{q}_c$ there is no phase transition and the Lyapunov exponent is found to be single valued. In the massive particle case, there is terminating temperature $\tilde{T}_t$ for Lyapunov exponent $\lambda$ at which it tends to zero.

In both the massless and massive particle cases we have studied the discontinuous change in the Lyapunov exponent $\lambda$. We have plotted the $\Delta \lambda  / \lambda_c$ vs $\tilde{T}_p / \tilde{T}_c$ curve and observe that when the Hayward AdS black hole undergoes small-large black hole phase transition, $\lambda$ moves from $\lambda_s$ to $\lambda_l$ with a non-zero change in the Lyapunov exponent $\Delta\lambda$. At the critical point $\Delta\lambda$ vanishes. The parameter $\Delta\lambda$ for Hayward AdS black hole, acts as an order parameter and the critical exponent near the critical point is $1/2$.

It will be interesting to study how this conjecture holds in different ensembles of black holes in different spacetime. We plan to do so in our future work.

\label{Section:Canonical}


\appendix*

\bibliographystyle{apsrev}
\end{document}